\documentclass[prb, twocolumn, superscriptaddress, showpacs]{revtex4}
\usepackage{tabularx,graphicx,subfigure}
\usepackage{dcolumn}
\usepackage{bm}
\usepackage{color}

\begin{document}

\title{First-Principles Simulations of Warm Dense Lithium Fluoride}

\author{K. P. Driver}
 \affiliation{Department of Earth and Planetary Science, University of California, Berkeley, California 94720, USA}
 \email{kdriver@berkeley.edu}
 \homepage{http://militzer.berkeley.edu/}

\author{B. Militzer}
 \affiliation{Department of Earth and Planetary Science, University of California, Berkeley, California 94720, USA}
 \affiliation{Department of Astronomy, University of California, Berkeley, California 94720, USA}

\date{\today}

\begin{abstract}
  We perform first-principles path integral Monte Carlo (PIMC) and
  density functional theory molecular dynamics (DFT-MD) calculations
  to explore warm dense matter states of LiF. Our simulations cover a
  wide density-temperature range of $2.08-15.70$~g$\,$cm$^{-3}$ and
  $10^4-10^9$~K.  Since PIMC and DFT-MD accurately treat effects of
  atomic shell structure, we find a pronounced compression maximum and
  a shoulder on the principal Hugoniot curve attributed to K-shell and
  L-shell ionization. The results provide a benchmark for widely-used
  EOS tables, such as SESAME, LEOS, and models. In addition, we
  compute pair-correlation functions that reveal an evolving plasma
  structure and ionization process that is driven by thermal and
  pressure ionization. Finally, we compute electronic density of
  states of liquid LiF from DFT-MD simulations and find that the
  electronic gap can remain open with increasing density and
  temperature to at least 15.7 g$~$cm$^{-3}$.

\end{abstract}



\maketitle

\section{INTRODUCTION}

Progress in our understanding of warm dense matter (WDM) relevant to
fusion energy and astrophysical phenomena relies on the development of
accurate techniques to determine the equation of state (EOS) of
materials across wide density-temperature regimes. The EOS provides
well-defined thermodynamic states that can be measured experimentally
in dynamic shock experiments and further used for hydrodynamic
modeling of experiments. The state of a shock in dynamic compression
experiments is often measured with impedance matching techniques via
an optically transparent interferometer window. While there are
several materials used for shock windows (quartz, diamond, MgO, etc),
LiF is frequently used because it has several favorable optical and
structural properties under
compression~\cite{Wise1986,Furnish1999,Huser2004}.

\begin{figure}[!htbp]
  \begin{center}
        \includegraphics*[width=8.6cm]{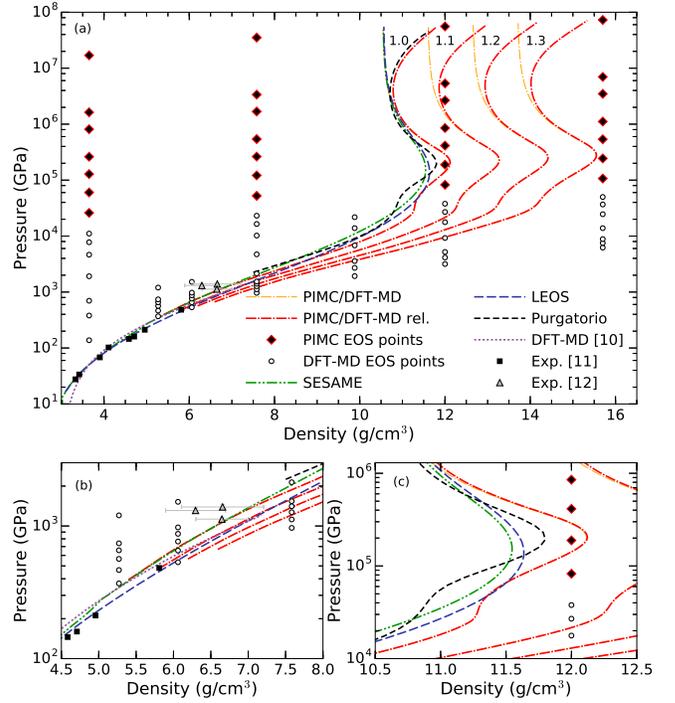}
  \end{center}

  \caption{(a) Comparison of PIMC/DFT-MD shock Hugoniot curves with
    SESAME-7271~\cite{SESAME1992,SESAMEhttp},
    LEOS-2240~\cite{More1988,Young1995,WhitleyLLNL} tables, and the
    Purgatorio (Lynx-2240)~\cite{Purgatorio2006} model, as well as
    previous DFT-MD~\cite{Clerouin2005} and
    experiments~\cite{Kormer1965,Hicks2003}, in P-$\rho$ space. The
    PIMC/DFT-MD Hugoniot curves are plotted for four initial
    densities, corresponding to 1--1.3-fold of ambient density.
    PIMC/DFT-MD and Purgatorio treat the quantum-mechanical shell
    structure of the ions and, thus, reveal a pronounced compression
    maximum and a shoulder due to K-shell and L-shell ionization
    effects. The SESAME and LEOS EOS tables are derived from models
    that do not explicitly treat shell effects. Plots (b) and (c) show
    a zoom in of the regions near the the experimental data and the
    compression maximum, respectively.}

  \label{fig:hugoniot1}
\end{figure}

Numerous shock
experiments~\cite{Kormer1965,Kormer1968,Hauver1970,gupta1975dislocation,gupta1975elastic,vorthman1982dislocations,tunison1986effect,whitlock1995orthogonal,Rigg1998,Rigg2001,Hicks2003,Jensen2007,LaLone2008,turneaure2009real,Ao2009,Fraizier2010,Fratanduono2011,Brown2013,dong2014compression,Rigg2014,liu2015sound,ZhiYu2015,Seagle2016}
have been performed to characterize the EOS, optical, and mechanical
properties of LiF in order to optimize its use as an interferometer
window. The shock Hugoniot curve has been measured up to 14
Mbar~\cite{Hicks2003}.  Experimental data has indicated that the large
LiF optical gap ($\sim$12 eV) decreases with compression to 800 GPa,
and, upon extrapolation, closes above 4000
GPa~\cite{Fratanduono2011}. In contrast, recent first-principles
simulations indicate that the optical gap should increase with density
to at least 500 GPa~\cite{Spataru2015,Sajid2013}.  Despite this
discrepancy, there is agreement between experiment and theory that the
refractive index increases linearly with density up to 800 GPa. In
addition, the EOS of LiF has been measured in diamond anvil
experiments~\cite{Pagannone1965,Yagi1978,Kim1976,Boehler1997,Liu2007}. From
these investigations, it is known that LiF has a high melting
temperature ($\sim$3000 K at 1 Mbar) and remains in the B1 structural
phase up to at least 1~Mbar, which makes LiF an excellent window
material in shock wave experiments. Additionally, ultrasonic
measurements~\cite{Briscoe1957,Jones1976,Hart1977} have been used to
measure elastic moduli.

Many theoretical investigations have also aimed to understand the
EOS~\cite{Clerouin2005,cherednikov2011atomistic,Smirnov2011,Sun2011},
electronic~\cite{norman2011excited,Sun2011} and
elastic~\cite{Zunger1977,Kunz1982,Doll1997},
thermodynamic~\cite{Belonoshko2000,stegailov2010stability,Smirnov2011,jones2016estimates}
transport~\cite{luo2016molecular}, and
optical~\cite{Clerouin2005,clerouin2006ab,Sun2011,Sajid2013,Spataru2015}
properties of LiF.  Most of these simulations focus on relatively
low-pressure and low-temperature regimes to help constrain the phase,
melt, and optical properties that are important for shock window
experiments. The highest temperature and pressure simulations at which
LiF has been studied so far, using density functional theory molecular
dynamics (DFT-MD) methods, were performed by Cl\'{e}rouin \emph{et
  al.}~\cite{Clerouin2005}, which predicted the EOS and shock Hugoniot
curve up to a density of 7 g$~$cm$^{-3}$ (14 Mbar) and a temperature
of 47,000 K~\cite{Clerouin2005}.

\begin{figure}[t]
  \begin{center}
        \includegraphics*[width=8.6cm]{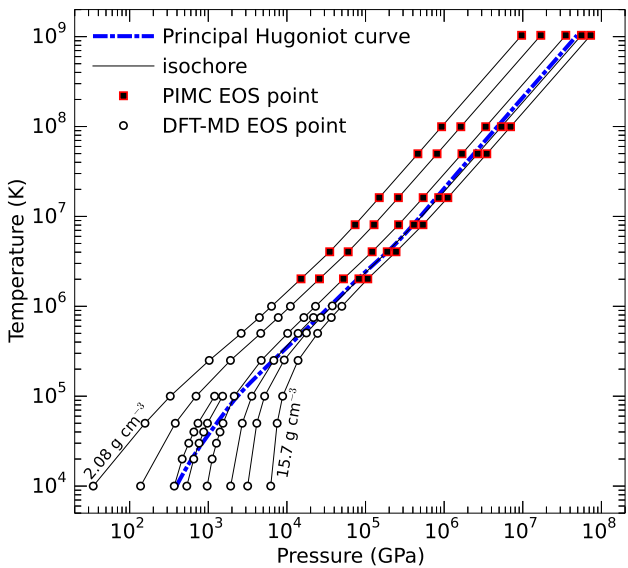}
  \end{center}

    \caption{Temperature-pressure conditions for the PIMC and DFT-MD
      calculations along four isochores corresponding to the densities
      of 2.082, 3.651, 7.582, and 15.701 g$\,$cm$^{-3}$. The blue,
      dash-dotted line shows the Hugoniot curve for an initial density
      of $\rho_0 = 2.635$ g$\,$cm$^{-3}$.}

  \label{fig:TvsP}
\end{figure}

\begin{figure}[t]
  \begin{center}
        \includegraphics*[width=8.6cm]{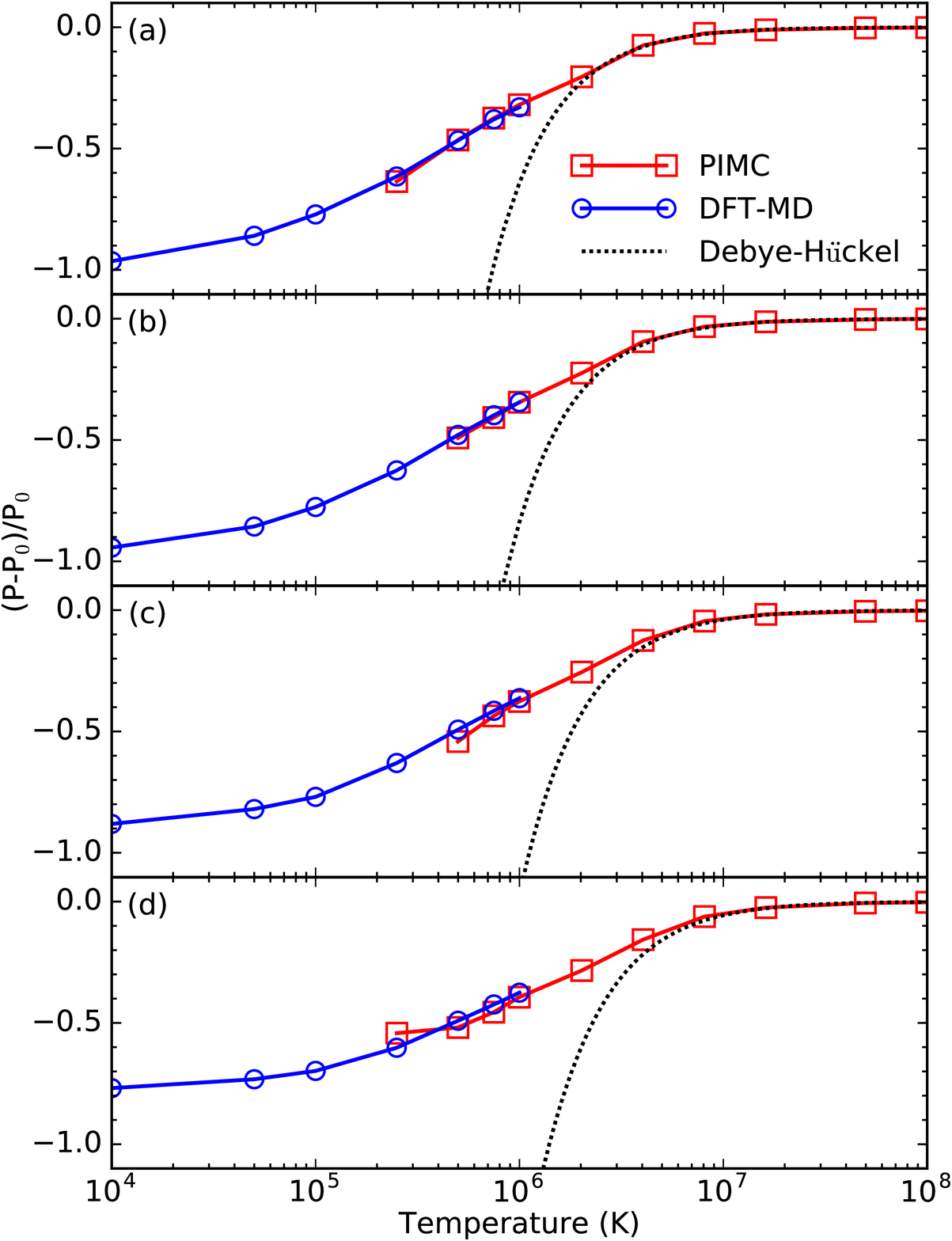}
  \end{center}

    \caption{LiF excess pressure, relative to the ideal Fermi gas,
      computed with PIMC, DFT-MD, and the Debye-H\"{u}ckel plasma
      model. The results are plotted for densities of (a) 2.082, (b)
      3.651, (c) 7.582, and (d) 15.701 g$\,$cm$^{-3}$ as a function of
      temperature.}

  \label{fig:PvsT}
\end{figure}

\begin{figure}[t]
  \begin{center}
        \includegraphics*[width=8.6cm]{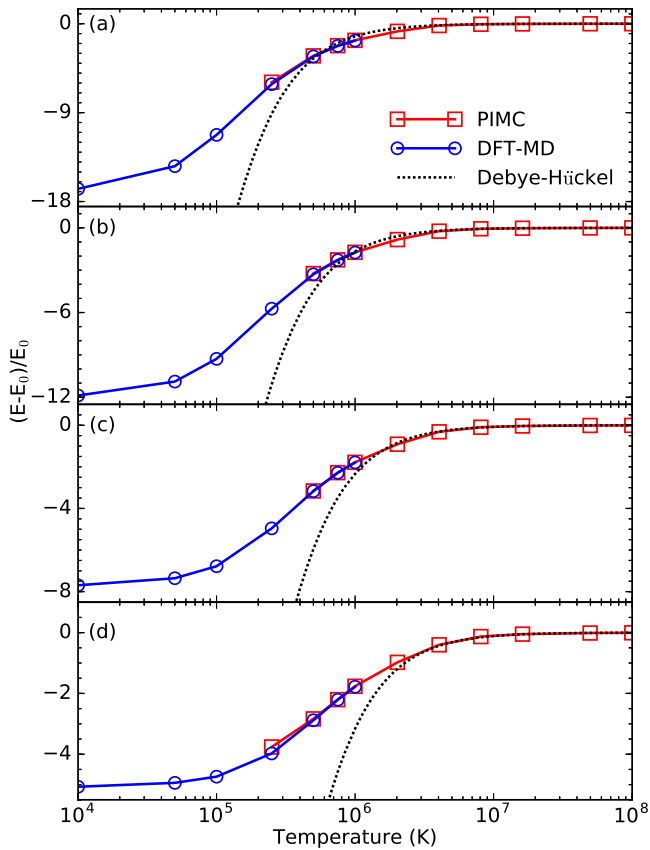}
  \end{center}

    \caption{LiF excess internal energy, relative to the ideal Fermi gas,
      computed with PIMC, DFT-MD, and the Debye-H\"{u}ckel plasma
      model. The results are plotted for densities of (a) 2.082, (b)
      3.651, (c) 7.582, and (d) 15.701 g$\,$cm$^{-3}$ as a function of
      temperature.}

  \label{fig:IEvsT}
\end{figure}

Because of the relevance of LiF for shock physics, it is desirable to
have a first-principles EOS derived for much higher temperature and
density conditions that span the the condensed matter, warm dense
matter, and plasma physics regimes as a reference for shock
experiments and hydrodynamic simulations. In recent works, we have
developed a first-principles framework to compute coherent EOSs across
a wide range of density-temperature regimes relevant to WDM by
combining results from state-of-the-art path integral Monte Carlo
(PIMC) and DFT-MD methods for first~\cite{Driver2012CarbonH2O}- and
second-row~\cite{Militzer2015Silicon} elements. In this paper, we
apply our PIMC and DFT-MD methods to compute the EOS and plasma
properties of LiF across a much larger density-temperature range than
has been studied in previous first-principles studies. We also study
the evolution of the plasma structure, ionization, and density of
states over the WDM regime. And, finally, we compare our PIMC/DFT-MD
shock Hugoniot curves with widely-used models and experiments.

The paper is organized as follows: Sec. II describes the simulation
methods. Section III provides the internal energy and pressure
EOS. Section IV discusses the shock Hugoniot curves. Section V
characterizes the plasma structure evolution and ionization processes
as a function of temperature and density via pair-correlation
functions. Section VI analyzes the electronic density of states as a
function of LiF density and pressure, and, finally, Sec. VII
summarizes our work.

\section{SIMULATION METHODS}

Rigorous discussions of the PIMC~\cite{Ce91,Ceperley1995,Ce96} and DFT
molecular dynamics (DFT-MD)~\cite{Car1985,Payne1992,Marx2009} methods
have been provided in previous works, and the details of our
simulations have been presented in our previous
publications~\cite{MC01,Mi09,Driver2012CarbonH2O,Benedict2014C,Driver2015Neon,
  Militzer2015Silicon,Driver2015Oxygen,Driver2016Nitrogen,Hu2016Si,Zhang2016Na,Zhang2017Na}. Here,
we summarize the methods and provide the simulation parameters
specific to LiF.

The general idea of our approach is to perform simulations along
isochores at high temperatures (T$\geq$1$\times$10$^6$~K) using PIMC
and at low temperatures (T$\leq$1$\times$10$^6$~K) using DFT-MD. We
show the two methods produce consistent results at overlapping
temperature regimes. The PIMC method samples the space of all quantum
particle paths to determine the thermal density matrix of the
many-body system. PIMC increases in efficiency with temperature (scaling
as 1/T) as quantum paths become shorter and more classical in
nature. In contrast, DFT-MD becomes increasingly inefficient with
increasing temperature, as the number of occupied bands increases
unfavorably with temperature (scaling roughly as $\sim$T$^{3/2}$). The
only uncontrolled approximation in PIMC is the use of the fixed-node
approximation, which restricts paths to avoid the well-known fermion
sign problem~\cite{Pierleoni1994}. We have shown the associated error
is small for relevant systems at high enough
temperatures~\cite{Ce91,Ce96,Driver2012CarbonH2O}. The main
approximation in DFT-MD is the use of an approximate
exchange-correlation (XC) functional, though at temperatures relevant to
WDM, error in the XC is small relative to the total energy, which is
the important quantity for EOS and Hugoniot
simulations~\cite{Karasiev2016}.

PIMC uses a small number of controlled approximations, whose errors
can be minimized by converging parameters, such as the time step and
system size. In simulations using free-particle nodes, we typically
use a time-step of 1/256 Ha$^{-1}$ for temperatures below
4$\times$10$^6$ K, where the total energy per atom is converged within
1\%. This results in using between 4 and 162 time slices for the
temperature range studied with PIMC (0.5$\times$10$^6$ to
1.034$\times$10$^9$~K).  Regarding finite size errors, we showed
simulations of 8- and 24-atom cubic cells provide internal energies
that agree within 1.0\% and pressures that agree within 0.5\% over the
relevant temperature range for PIMC
(T$>$1$\times$10$^6$~K)~\cite{Driver2015Neon}. Our results for the
internal energy and pressure typically have statistical errors of
0.3\% or less.

\begin{figure}
  \begin{center}
        \includegraphics*[width=8.6cm]{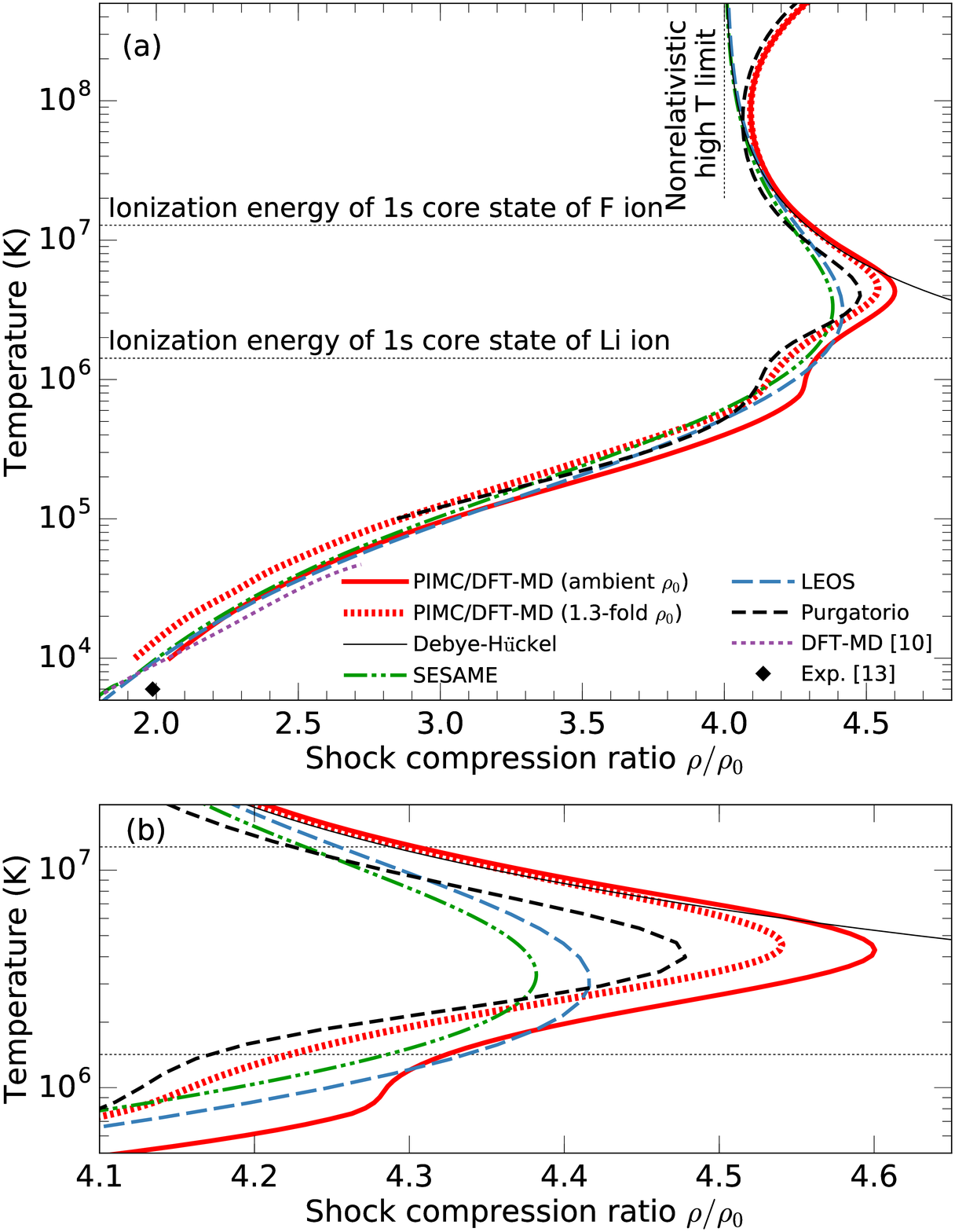}
  \end{center}

  \caption{(a) Comparison of PIMC/DFT-MD shock Hugoniot curves with
    SESAME-7271~\cite{SESAME1992,SESAMEhttp},
    LEOS-2240~\cite{More1988,Young1995,WhitleyLLNL} tables, and the
    Purgatorio (Lynx-2240)~\cite{Purgatorio2006} and
    Debye-H{\"u}ckel~\cite{DebyeHuckel} models, as well as previous
    DFT-MD~\cite{Clerouin2005} and experiments~\cite{Kormer1968}, in
    T-$\rho/\rho_0$ space. The PIMC/DFT-MD Hugoniot curves are plotted
    for two initial, pre-compressed density states, corresponding to
    1-fold and 1.3-fold of ambient. As in Fig.~\ref{fig:hugoniot1},
    PIMC/DFT-MD and Purgatorio predict shell structure effects along
    the Hugoniot, while SESAME and LEOS predict the overall behavior
    without shell effects. Plot (b) shows a zoom in of the compression
    maximum region.}

  \label{fig:hugoniot2}
\end{figure}

\begin{figure}
  \begin{center}
        \includegraphics*[width=8.6cm]{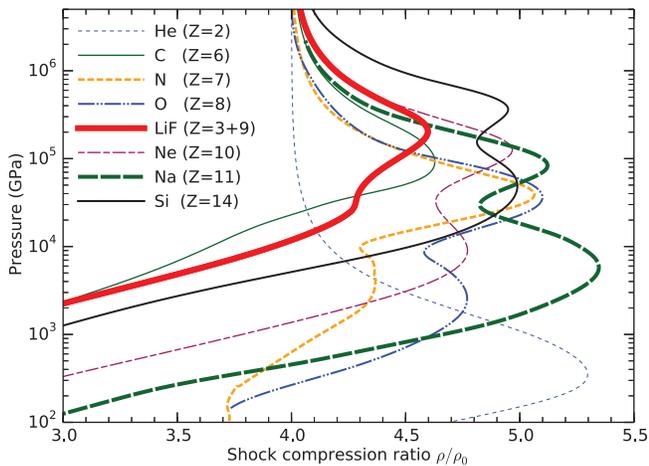}
  \end{center}

    \caption{Comparison of shock Hugoniot curves in P-$\rho/\rho_0$
      space for various materials initialized at ambient or
      experimental densities. The initial densities (in g~cm$^{-3}$)
      are He: 0.124, C: 2.253, N 0.807, O: 0.667, LiF: 2.635, Ne:
      1.507, and Si: 2.329. }

  \label{fig:HugComparison}
\end{figure}

\begin{figure}
  \begin{center}
        \includegraphics*[width=8.6cm]{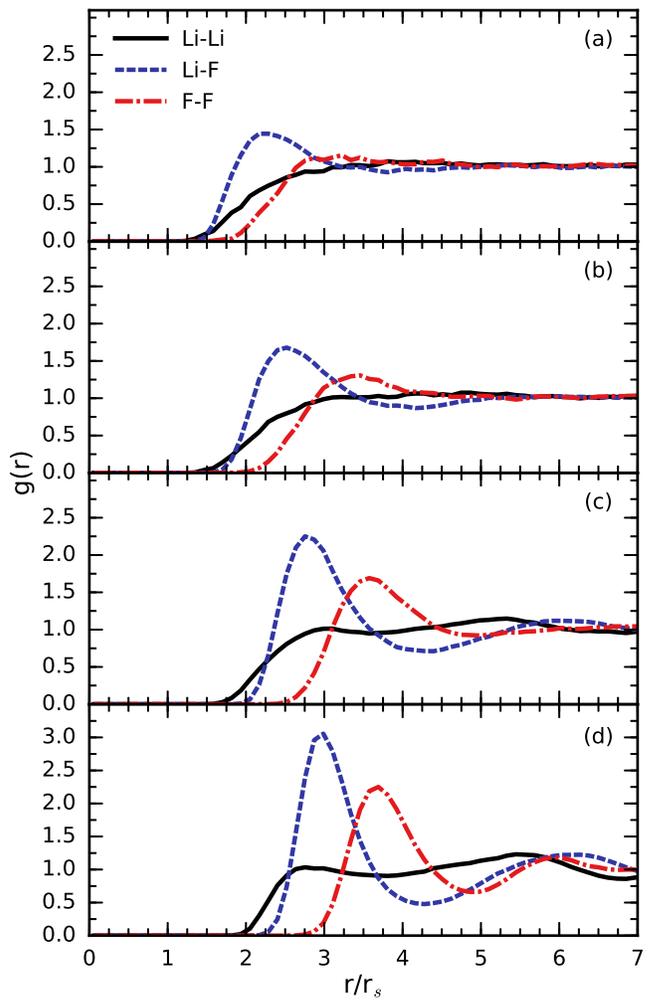}
  \end{center}

    \caption{Nuclear pair correlation functions computed with DFT-MD
      simulations of LiF liquid at a fixed temperature of
      2$\times$10$^4$ K.  Functions are compared for densities of (a)
      2.082, (b) 3.651, (c) 7.582, and (d) 15.701 g~cm$^{-3}$ (64-atom
      simulation cells).}

  \label{GofRLowT}
\end{figure}

\begin{figure}[t]
  \begin{center}
        \includegraphics*[width=8.6cm]{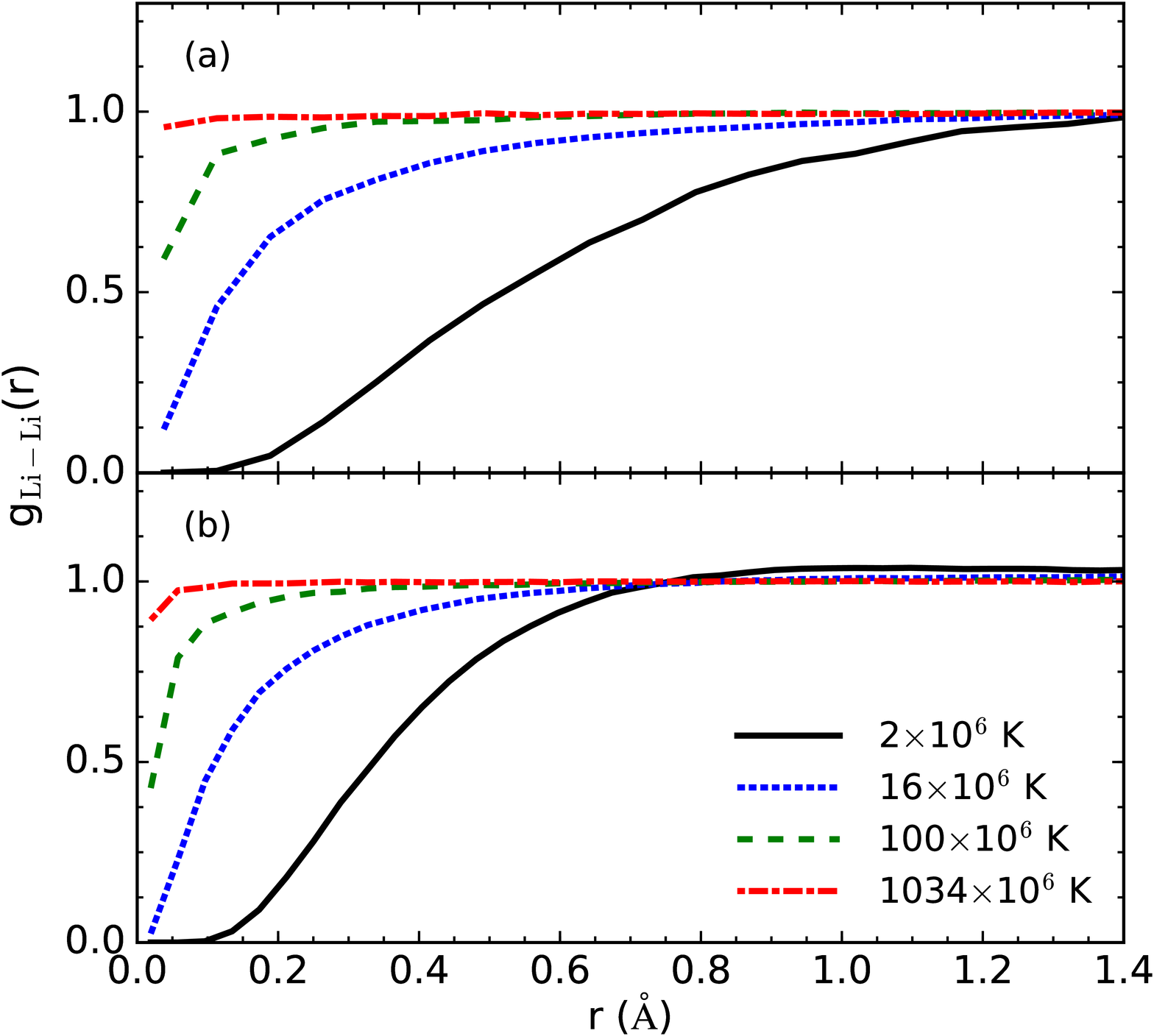}
  \end{center}

    \caption{Pair-correlation functions of Li nuclei computed with
      PIMC simulations of LiF over a wide range of
      temperatures. Functions are compared for densities of (a) 2.082
      and (b) 15.701 g~cm$^{-3}$ (8-atom simulation cells).}

  \label{fig:gofrlili}
\end{figure}

\begin{figure}[t]
  \begin{center}
        \includegraphics*[width=8.6cm]{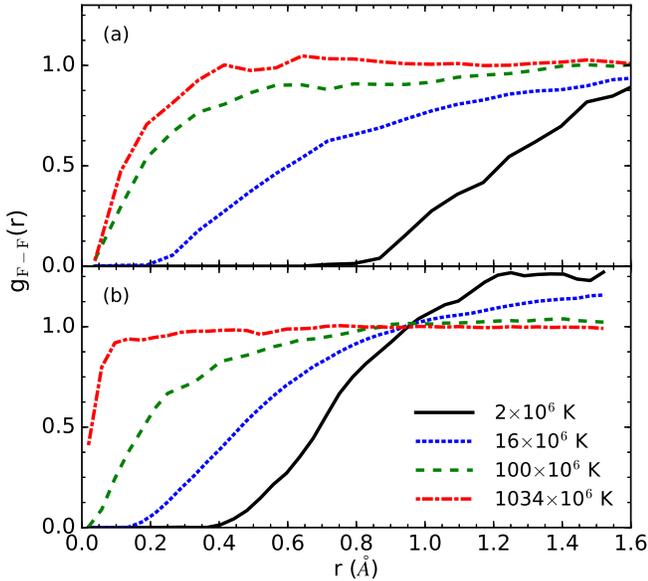}
  \end{center}

    \caption{Pair-correlation functions of F nuclei computed with PIMC
      simulations of LiF over a wide range of temperatures. Functions
      are compared for densities of (a) 2.082 and (b) 15.701
      g~cm$^{-3}$ (8-atom simulation cells).}

  \label{fig:gofrflfl}
\end{figure}

We employ standard Kohn-Sham DFT-MD simulation techniques for our low
temperature (T $\leq$ 1$\times$10$^6$ K) calculations of warm dense
LiF.  Simulations are performed with the Vienna \emph{Ab initio}
Simulation Package (VASP)~\cite{VASP} using the projector
augmented-wave (PAW) method~\cite{PAW,Kresse1999PAW}, and a NVT
ensemble, regulated with a Nos\'{e} thermostat. Exchange-correlation
effects are described using the Perdew-Burke-Ernzerhof~\cite{PBE}
generalized gradient approximation. Electronic wave functions are
expanded in a plane-wave basis with a energy cut-off as high as 4000
eV in order to converge total energy.  Size convergence tests up to a
64-atom simulation cell at temperatures of 10,000 K and above indicate
that internal energies are converged to better than 0.1\% and
pressures are converged to better than 0.6\%. We find, at temperatures
above 250,000 K, 8-atom supercell results are sufficient for both
energy and pressure since the kinetic energy far outweighs the
interaction energy at such high
temperatures~\cite{Driver2015Neon}. The number of bands in each
calculation were selected such that orbitals with occupation as low as
10$^{-4}$ were included, which requires up to 7,500 bands in a 24-atom
cell at 1$\times$10$^6$ K. All simulations are performed at the
$\Gamma$-point of the Brillouin zone, which is sufficient for high
temperature fluids, converging total energy to better than 0.01\%
compared to a grid of k-points.

\section{EQUATION OF STATE RESULTS}

In this section, we report our combined PIMC and DFT-MD EOS results
for the liquid, WDM, and plasma regimes at several densities in the
range of 2.082--15.701 g$~$cm$^{-3}$ and temperatures ranging from
10$^4$--10$^9$ K. The full-range of our EOS data is shown in
pressure-density space in Fig.~\ref{fig:hugoniot1} and in
temperature-pressure space in Fig.~\ref{fig:TvsP}. These two figures
will be discussed more thoroughly in Section VI. The Supplemental
Material~\cite{SuppMat} provides a table of our full EOS data set. In
order to put the PAW-PBE pseudopotential energies on the same scale as
all-electron calculations, we shifted all of our VASP DFT-MD energies
by -107.061113 Ha/LiF. This shift was determined by performing
isolated, all-electron atomic calculations with the OPIUM
code~\cite{OPIUM} and corresponding isolated-atom calculations using
the appropriate pseudopotential in VASP.

\begin{figure}[!htbp]
  \begin{center}
        \includegraphics*[width=8.6cm]{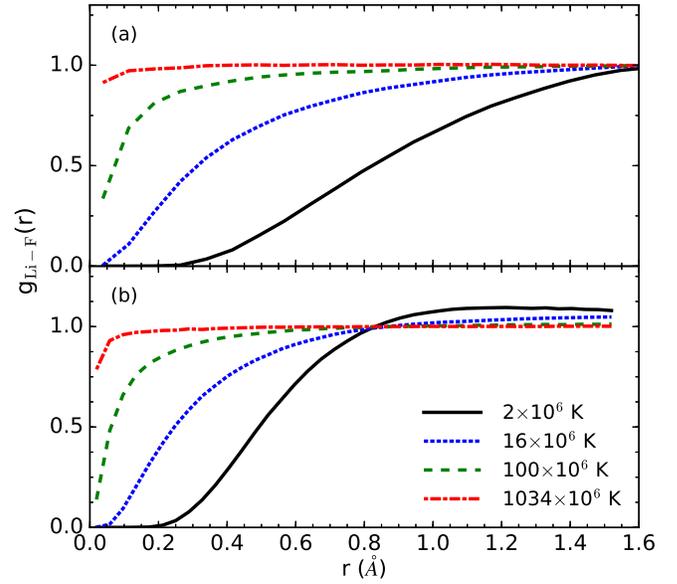}
  \end{center}

    \caption{Pair-correlation functions of Li-F nuclei computed with
      PIMC simulations of LiF over a wide range of
      temperatures. Functions are compared for densities of (a) 2.082
      and (b) 15.701 g~cm$^{-3}$ (8-atom simulation cells).}

  \label{fig:gofrlifl}
\end{figure}

In order to analyze the behavior of our EOS data, Figs.~\ref{fig:PvsT}
and~\ref{fig:IEvsT} compare pressure and internal energy,
respectively, along four isochores from PIMC, DFT-MD, and the
classical Debye-H{\"u}ckel plasma model~\cite{DebyeHuckel} as a
function of temperature. The pressures, P, and internal energies, E,
are plotted relative to a fully ionized Fermi gas of electrons and
ions with pressure, P$_0$, and internal energy, E$_0$, in order to
compare only the excess pressure and internal energy contributions
that result from particle interactions. With increasing temperature,
the pressure and internal contributions due to interactions gradually
decrease from the strongly-interacting condensed matter regime, where
bound states dominate, to the weakly-interacting, fully-ionized plasma
regime, where agreement is found with the Debye-H{\"u}ckel model. As
one expects, the classical Debye-H{\"u}ckel model becomes inadequate
for lower temperatures (T$<$5$\times$10$^6$~K) since it fails to treat
bound electronic states. While the range of temperatures over which
PIMC EOS data is needed to fill the temperature gap between DFT-MD and
Debye-H{\"u}ckel (roughly 2$-$5$\times$10$^6$~K) is relatively small
compared to the entire temperature range of the high energy density
physics regime, this temperature range encompasses the important
process of K-shell ionization, which is precisely where the full rigor
of PIMC is needed to acquire an accurate EOS table.

The two figures together provide a coherent EOS over wide
density-temperature range for LiF due to the fact that PIMC and DFT-MD
provide consistent, overlapping results, with a maximum difference of
3\% in the pressure and 3.6\% ($\sim$1.7 Ha/LiF) in the internal
energy at 1$\times$10$^6$ K. Furthermore, this agreement between PIMC
and DFT-MD provides validation for the use of zero-temperature
exchange correlation functionals in WDM applications and the use of
free-particle nodes in PIMC. However, as noted by Karasiev \emph{et
  al.}~\cite{Karasiev2016}, this may only be true when the total
energy is large relative to the exchange-correlation
energy. Finite-temperature exchange-correlation contributions were
predicted to be significant for other properties, such as conductivity
at low densities.  At lower temperatures, PIMC results become
inconsistent with DFT-MD results because the free-particle nodal
approximation in PIMC simulations is no longer appropriate.

\begin{figure}[!htbp]
  \begin{center}
        \includegraphics*[width=8.6cm]{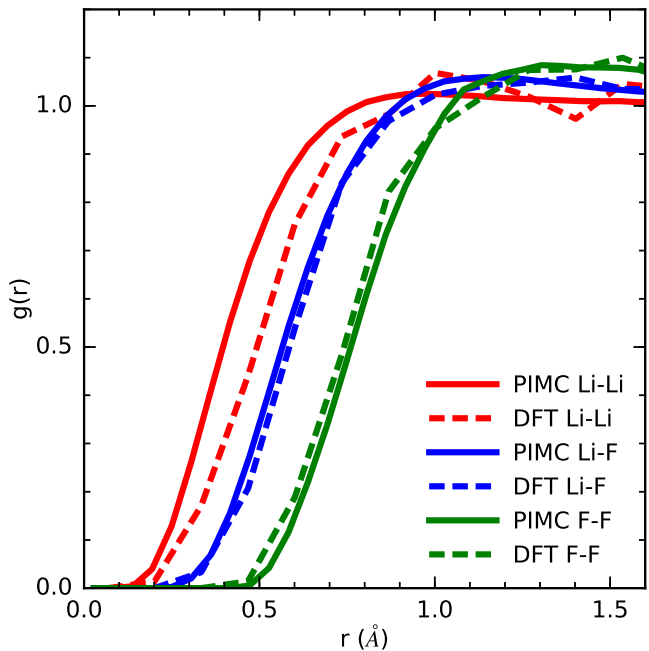}
  \end{center}

    \caption{Comparison of nuclear pair-correlation functions computed
      with PIMC and DFT-MD for LiF at a temperature of 1$\times$10$^6$
      K and a density of 15.701 g$\,$cm$^{-3}$ (24-atom simulation
      cells).}

  \label{fig:gofrlifl1millionK}
\end{figure}

\section{SHOCK COMPRESSION}

Dynamic shock compression experiments allow one to directly measure
the equation of state and other physical properties of hot, dense
fluids.  Such experiments are often used to determine the principal
Hugoniot curve, which is the locus of final states that can be
obtained from different shock velocities.  Density functional theory
has been validated by experiments as an accurate tool for predicting
the shock compression of a variety of different
materials~\cite{Root2010,Wang2010,Mattsson2014}.

During a shock wave experiment, a material whose initial
state is characterized by an internal energy, pressure, and volume,
($E_0,P_0,V_0$), will change to a final state denoted by $(E,P,V)$
while conserving mass, momentum, and energy. This leads to the
Rankine-Hugoniot relation~\cite{Ze66},

\begin{equation}
(E-E_0) + \frac{1}{2} (P+P_0)(V-V_0) = 0.
\label{hug}
\end{equation}

Here, we solve this equation for our computed first-principles EOS
data set, which is reported in the Supplemental
Material~\cite{SuppMat}. We obtain a continuous Hugoniot curve by
interpolating the EOS data with a rectangular bivariate spline as a
function of $\rho$ and $T$. We have compared several different spline
algorithms and find the differences are negligible given that
reasonable choices are made for the isochore densities with respect to
Hugoniot features. In order to obtain the principal Hugoniot curve, we
used initial conditions based on the energy and pressure of ambient,
solid LiF in the B1 phase computed with static DFT (P$_0$ = 3.323 GPa,
E$_0$ = $-$107.417375 Ha/LiF, V$_0=$16.346636 $\rm \AA^3/LiF$,
$\rho_0=$2.635 g$\,$cm$^{-3}$). The resulting Hugoniot curve has been
plotted in $P$-$\rho$ space in Fig.~\ref{fig:hugoniot1}, in $T$-$P$
space in Fig.~\ref{fig:TvsP}, and in $T$-$\rho/\rho_0$ space in
Fig.~\ref{fig:hugoniot2}.

Samples in shock wave experiments may be precompressed inside of a
diamond anvil cell before the shock wave is launched in order to reach
much higher final densities than are possible with a sample at ambient
conditions~\cite{Militzer2007,Jeanloz2007PNAS}. This technique allows shock wave
experiments to probe a density-temperature regimes consistent with
planetary and stellar interiors. Therefore, we repeat our Hugoniot
calculation starting with initial densities of 1.1- to 1.3-fold of the
ambient density.


Fig.~\ref{fig:hugoniot2} shows the temperature along the Hugoniot
curve as a function of the shock-compression ratio for the principal
Hugoniot curve and a curve corresponding to 1.3-fold
precompression. Consistent with our studies of other elements, we find
that an increase in the initial density leads to a slight reduction in
the shock compression ratio because particles interact more strongly
at higher density. In the high-temperature limit, all curves converge
to a compression ratio of 4, which is the value of a nonrelativistic,
ideal gas.  We also show the magnitude of the relativistic correction
to the Hugoniot in the high-temperature limit. The shock compression
and structure along the Hugoniot is determined by the excitation of
internal degrees of freedom, such as dissociation and ionization
processes, which increases the compression, and, in addition, the
interaction effects, which decrease the compression~\cite{Mi06}.

In the structure of the principal Hugoniot curve, we identify a
pronounced compression maximum at high temperature and a shoulder at
lower temperature, which correspond to the ionization of the K-shell
and L-shell in LiF.  The lower-temperature shoulder on the principal
Hugoniot curve occurs near a compression ratio of $\rho/\rho_0$ = 4.28
and a temperature of 9.00$\times$10$^5$~K (77.56 eV), which
corresponds to the K-shell ionization of lithium and the L-shell
ionization of fluorine. The K-shell ionization energies of lithium are
75.64 eV (8.78$\times$10$^5$~K) and 122.45 eV
(1.42$\times$10$^6$~K)~\cite{NIST}. The 2p state of lithium is already
ionized by a temperature of 5.39 eV (6$\times$10$^4$~K).  The L-shell
ionization energies of fluorine range from 17.42--185.19 eV
(0.2-2.1$\times$10$^6$~K). The higher-temperature compression maximum
at $\rho/\rho_0$ = 4.54 on the principal Hugoniot curve occurs at
temperature of 4.53$\times$10$^6$~K (365.29 eV), which corresponds to
the K-shell ionization in fluorine. The K-shell ionization energies of
fluorine are 953.89 and 1103.12 eV (11.1 and
12.8$\times$10$^6$~K). This is consistent with the ionization process
we observe in Figs.~\ref{fig:NofRLi} and \ref{fig:NofRF}, where charge
density around the nuclei is reduced over the range of
1-8$\times$10$^6$~K. Propagating errors from our equation of state
data into the Hugoniot curve shows that the statistical uncertainty
in the density along the Hugoniot is at most 4\% and, the statistical
error in the pressure along the Hugoniot is at most 3\%.

Figure~\ref{fig:hugoniot1} shows the principal and pre-compressed
Hugoniot curves in $P-\rho$ space. Starting from ambient density
(2.635 g$~$cm$^{-3}$), the compression maximum occurs at a density of
12.113 g$~$cm$^{-3}$ (4.596-fold compression) and a pressure of
1.988$\times$10$^5$ GPa. Starting with a precompressed density of
1.3-fold of ambient (3.426 g$~$cm$^{-3}$), the compression maximum
occurs at a density of 15.537 g$~$cm$^{−3}$ (5.897-fold compression)
and a pressure of 2.704$\times$10$^5$ GPa. Alternatively, higher
densities can be reached with multi-shock
experiments~\cite{Smith2014Ramp}.

In both Fig.~\ref{fig:hugoniot1} and Fig.~\ref{fig:hugoniot2}, we
compare our PIMC principal Hugoniot curve with several, widely-used
EOS tables and models, such as SESAME (Table
7271)~\cite{SESAME1992,SESAMEhttp}, LEOS (Table
2240)~\cite{More1988,Young1995,WhitleyLLNL}, Purgatorio
(Lynx-2240)~\cite{Purgatorio2006}, and
Debye-H{\"u}ckel~\cite{DebyeHuckel}. The SESAME and LEOS models are
largely based on variations of the Thomas-Fermi model, which treats
electrons in an ion-sphere as a non-uniform electron gas, neglecting
quantum-mechanical shell structure of the Li and F nuclei. Therefore,
we see that, while the SESAME and LEOS Hugoniot curves provide good
overall agreement with PIMC in this case, they do not exhibit any
compression maximum related to shell structure. On the other hand, the
DFT-based, average-atom Purgatorio (Lynx) model does compute the shell
structure for an average of multiple ionic states. Thus, Purgatorio
predicts the correct ionization features, a shoulder and a
well-defined compression maximum, along the principal Hugoniot curve
in good agreement with PIMC. However, overall the Purgatorio Hugoniot
curve is slightly less compressible than the PIMC
prediction. Remarkably, Purgatorio achieves this level of accuracy
while being 100-1000$\times$ more efficient than Kohn-Sham DFT-MD and
PIMC. We note that the classical Debye-H{\"u}ckel model is excellent
agreement with PIMC for temperatures above 5$\times$10$^6$~K. This
means that for LiF, PIMC is only needed to fill a relatively small gap
in temperature (2$\times$10$^6$~K and 5$\times$10$^6$~K) between DFT-MD
and Debye-H{\"u}ckel EOS data, which encompasses the K-shell
compression peak.

We also compare with experimental data in Fig.~\ref{fig:hugoniot1} and
Fig.~\ref{fig:hugoniot2}, which is available for low temperatures and
pressures. At lowest temperatures in Fig.~\ref{fig:hugoniot2}, the
Hugoniot curves of all models lie slightly above the shock melting
measurement by Kormer~\cite{Kormer1968}. In Fig.~\ref{fig:hugoniot1},
all models agree reasonably well with the experimental liquid shock
data of Kormer \emph{et al.}~\cite{Kormer1965}, given there may be
slight differences in initial shock conditions.  The lowest-pressure
experimental data from Hicks \emph{et al.}~\cite{Hicks2003} lies about
250 GPa above previous DFT-MD~\cite{Clerouin2005} and LEOS results,
while SESAME and the DFT-MD calculations presented here pass through
the lowest pressure data point.

Finally, Fig.~\ref{fig:HugComparison} compares our LiF Hugoniot curve
in P-$\rho/\rho_0$ space with our previous first-principles Hugoniot
curves for other first- and second-row
materials~\cite{Driver2016HEDP}, He~\cite{Mi09},
C~\cite{Driver2012CarbonH2O,Benedict2014C},
N~\cite{Driver2016Nitrogen}, O~\cite{Driver2015Oxygen},
Ne~\cite{Driver2015Neon}, Na~\cite{Zhang2016Na,Zhang2017Na}, and
Si~\cite{Militzer2015Silicon,Hu2016Si}.  For each Hugoniot curve,
DFT-MD data is plotted for T$<$1$\times$10$^6$ K, and PIMC results are
plotted for higher temperatures. Each Hugoniot curve exhibits at least
one distinct shock-compression maximum corresponding to K or L shell
ionization.  The maximum compression ratio reached in each case is
largely determined be the initial density of the system due to
interaction effects. Helium has the lowest initial density and highest
maximum compression ratio, while LiF has the highest initial density
and lowest maximum compression ratio. The pressure (and temperature)
at the compression maximum scales with roughly with the binding
energy, Z$^2$, which means a higher pressure (or temperature) is
needed to reach the regime of ionization. Therefore, as a general
trend, as Z increases, the compression peak temperatures increase. In
the high-temperature limit, all curves converge to a compression ratio
of 4, which is the value of a nonrelativistic, ideal gas.

\section{PAIR-CORRELATION FUNCTIONS}

In this section, we provide a discussion of the temperature and
density dependence of pair-correlation, g(r), functions and ionization
processes in warm dense LiF. The radial pair correlation function is
defined as

\begin{equation}
g(r) = \frac{V}{ 4 \pi r^2 N^2} \left< \sum_{j > i}\delta(r-r_{ij}) \right>,
\end{equation}
where N is the total number of particles, V is the cell volume, and r
is the distance from the ith reference particle.

Fig.~\ref{GofRLowT} shows Li-Li, Li-F, and F-F pair-correlation curves
at a fixed, low temperature of 2$\times$10$^4$~K computed with DFT-MD
simulations in 64-atom cells at four densities. We plot the g(r)
functions as a function of r/r$_s$, where r$_s$=$(3/(4 \pi
n_e))^{1/3}$ and n$_e$ is the electron number density of LiF, in order
to clearly differentiate between correlation- and density-driven
changes. The results we find are in a good agreement with trends found
by Cl\'{e}rouin \emph{et al.}~\cite{Clerouin2005}, but investigated
for a larger density range. As in most fluids, LiF becomes more
structured with increasing density. We also note that higher
temperatures always result in less structured fluid at each
density. Compared to the other g(r) functions, there is less structure
in Li-Li curves, which implies those nuclei interact weakly. By
examining the mean square displacements as a function of time, we also
find the lithium atoms diffuse much faster due to their lighter
mass. The fluorine atoms exhibit increasing strong correlations with
density, preserving an ionic fluid structure, as seen in the F-F g(r)
peak. Furthermore, the lithium atoms remain strongly correlated with
ionic fluorine structure due to Coulomb interactions, as seen in the
Li-F g(r) peak.

\begin{figure}[!htbp]
  \begin{center}
        \includegraphics*[width=8.6cm]{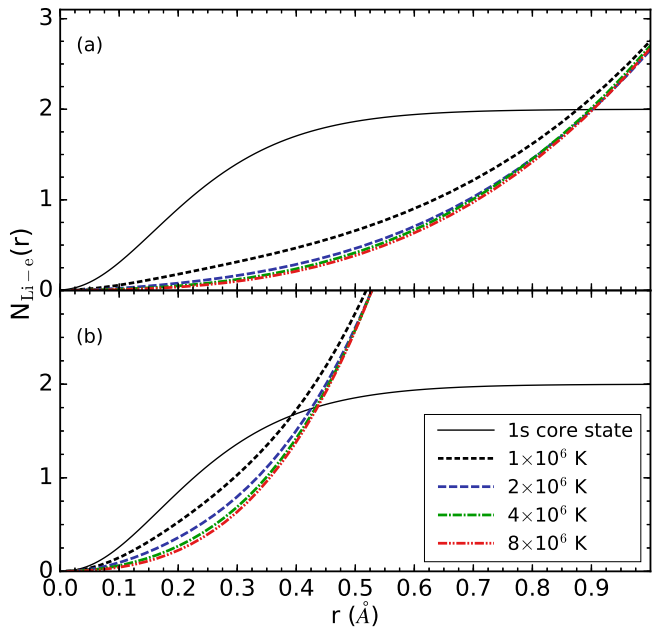}
  \end{center}

    \caption{Number of electrons contained in a sphere of radius, $r$,
      around a lithium nucleus in LiF plasma. PIMC data at two
      densities of (a) 2.082 and (b) 15.701 g~cm$^{-3}$ and four
      temperatures is compared with the doubly occupied lithium 1s
      core ground state (8-atom simulation cells).}

  \label{fig:NofRLi}
\end{figure}

Figures~\ref{fig:gofrlili}, \ref{fig:gofrflfl}, and \ref{fig:gofrlifl}
show ion-ion $g(r)$ curves for Li-Li, F-F, and Li-F pairs in LiF
plasmas, respectively. The g(r) functions were computed with PIMC at
temperatures relevant to WDM for four densities. We first note that
g(r) curves corresponding to heavier ions are systematically found
further apart than lighter ions due to stronger Coulomb repulsion and
Pauli exclusion from bound electrons. In each case, the atoms are kept
farthest apart at low temperatures.  As temperature increases, kinetic
energy of the nuclei increases, leading to stronger collisions and
making it more likely to find them at close range. At the same time,
the atoms become increasingly ionized, which gradually reduces the
Pauli repulsion, while increasing the ionic Coulomb repulsion. At the
highest temperatures, the system approaches the Debye-H{\"u}ckel
limit, behaving like a weakly correlated system of screened Coulomb
charges. As density increases, the likelihood of finding two nuclei at
close range rises only slightly.

In order to show that PIMC and DFT-MD predict similar plasma
structures, Fig.~\ref{fig:gofrlifl1millionK} compares Li-Li, Li-F, and
F-F g(r) curves using both methods at 1$\times$10$^6$~K at our highest
isochore density of 15.701 g$~$cm$^{-3}$ in 24-atom simulation
cells. The fact that the PIMC and DFT-MD g(r) curves nearly overlap
indicates that both methods predict a consistent ionic plasma
structure in addition to a consistent EOS. There are some small
differences in the Li-Li g(r) DFT-MD and PIMC curves likely due to
frozen-core pseudopotentials and exchange correlation effects.

\begin{figure}[!htbp]
  \begin{center}
        \includegraphics*[width=8.6cm]{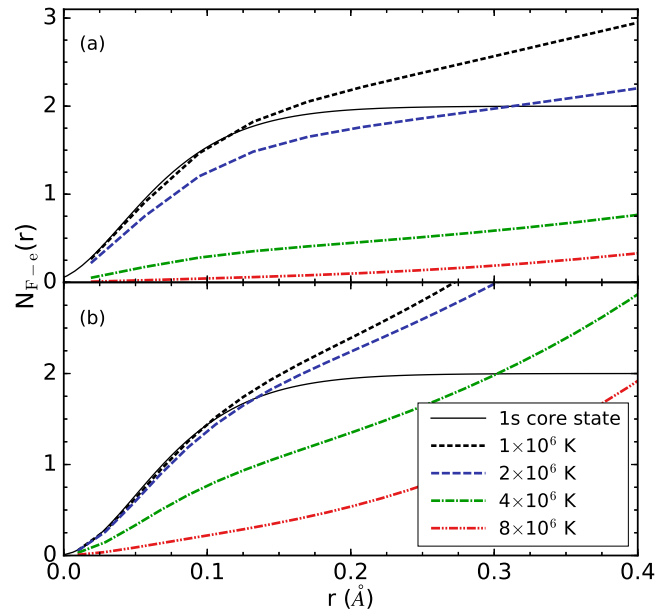}
  \end{center}

    \caption{Number of electrons contained in a sphere of radius, $r$,
      around a fluorine nucleus in LiF plasma. PIMC data at two
      densities of (a) 2.082 and (b) 15.701 g~cm$^{-3}$ and four
      temperatures is compared with the doubly occupied fluorine 1s
      core ground state (8-atom simulation cells).}

  \label{fig:NofRF}
\end{figure}

Figs.~\ref{fig:NofRLi} and \ref{fig:NofRF} show the integral of the
nucleus-electron pair correlation function, $N(r)$, for Li-e and F-e
in LiF plasma, respectively, as a function of temperature and
density. N(r) represents the average number of electrons within a
sphere of radius $r$ around a given nucleus. N(r) is given by the
formula
\begin{equation}
N(r) = \left< \frac{1}{N_I} \sum_{e,I}
\theta(r-\left|\vec{r}_e-\vec{r}_I \right|) \right>,
\end{equation}
where the sum includes all electron-ion pairs and $\theta$ represents
the Heaviside function. 

\begin{figure}[!htbp]
  \begin{center}
        \includegraphics*[width=8.6cm]{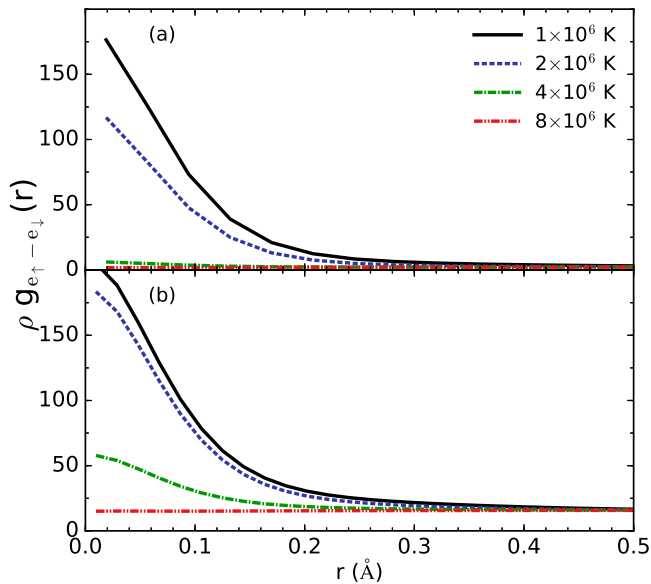}
  \end{center}

    \caption{The electron-electron pair-correlation functions for
      electrons with opposite spins in PIMC calculations of LiF
      plasma. Results are compared for densities of (a) 2.082 and (b)
      15.701 g~cm$^{-3}$ at four temperatures (8-atom cells).}

  \label{fig:gofreeopp}
\end{figure}

\begin{figure}[t]
  \begin{center}
        \includegraphics*[width=8.6cm]{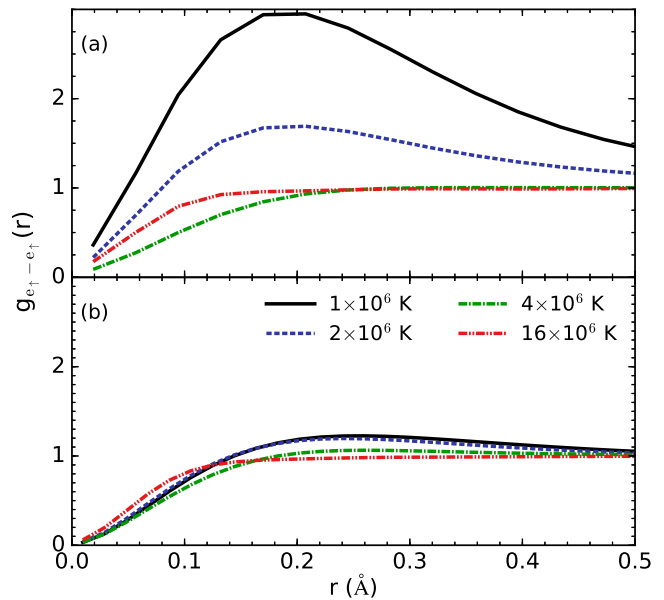}
  \end{center}

    \caption{The electron-electron pair-correlation functions for
      electrons with parallel spins in PIMC calculations of LiF
      plasma. Results are compared for densities of (a) 2.082 and (b)
      15.701 g~cm$^{-3}$ at four temperatures. (8-atom cells).}

  \label{fig:gofreesame}
\end{figure}

In each figure, we compare our PIMC N(r) curves with the 1s
ground-state of a corresponding isolated Li or F atom to gauge the
extent of ionization.  It is clear from Fig.~\ref{fig:NofRLi} that
the lithium ion is almost fully ionized for all temperatures and
densities considered. While there are some partially bound states
remaining in the lithium ions at a temperature of 1$\times$10$^6$~K,
by 8$\times$10$^6$~K the Li atoms have been fully ionized.  In
contrast, from Fig.~\ref{fig:NofRF} it is clear that the higher-Z,
fluorine ion still has bound 1s electrons at 1$\times$10$^6$~K due to
a higher binding energy. As temperature increases, the K-shell of the
fluorine ions gradually becomes more ionized, causing N(r) to
decrease. As density increases, it is apparent that higher
temperatures are required to fully ionize the fluorine ion. Thus, we
observe that the 1s ionization fraction decreases with density, which
indicates that pressure ionization of the fluorine K-shell is absent,
as we have observed for other first- and second-row elements in our
previous work~\cite{Driver2015Oxygen,Driver2016HEDP}.

Fig.~\ref{fig:gofreeopp} shows electron-electron pair correlations in
LiF plasma for electrons having opposite spins. The functions are
multiplied by the mass number density $\rho$, so that the integral
under the curves is proportional to the number of electrons. The
electrons are most highly correlated at low temperatures, which
reflects that multiple electrons occupy bound states around a given
nucleus. As temperature increases, electrons are thermally excited,
decreasing the correlation among each other. The positive correlation
at short distances increases with density, consistent with a lower
ionization fraction seen in our N(r) plots.

Fig.~\ref{fig:gofreesame} shows electron-electron pair correlations in
LiF plasma for electrons with parallel spins.  The positive
correlation at intermediate distances (r $\approx$ 0.2 $\rm \AA$)
reflects that different electrons with parallel spins are bound to a
given nucleus. For short separations, electrons strongly repel due to
Pauli exclusion and the functions decay to zero.  As density
increases, the peak at intermediate distances decreases and clearly
shows the effect of pressure ionization of the L shell. Pressure
ionization is expected for L-shell orbitals because they are much
larger than the K-shell orbitals and are therefore subject to Pauli
exchange with nearby nuclei. As temperature increases, electrons
become less bound, which also causes the correlation to become more
like an ideal fluid.

\section{ELECTRONIC DENSITY OF STATES}
In this section, we report DFT-MD results for the electronic density
of states (DOS) as a function of temperature and density in the liquid
and plasma states of LiF. This analysis provides further insight into
the temperature-density evolution of ionization effects and the band
gap. All DOS curves were computed with 64-atom simulation cells. At
this cell size and temperature range (1$\times$10$^4$--5$\times$10$^5$
K), we found a single k-point provides sufficiently converged DOS
results. Smooth curves were obtained by averaging over a MD simulation
and applying a Gaussian smearing of 0.5~eV to the band
energies. Furthermore, the eigenvalues of each snapshot were shifted
so that the Fermi energies, E$_{\rm F}$, align at zero, and the
integral of the occupied DOS is normalized to 1.

\begin{figure}[t]
  \begin{center}
        \includegraphics*[width=8.6cm]{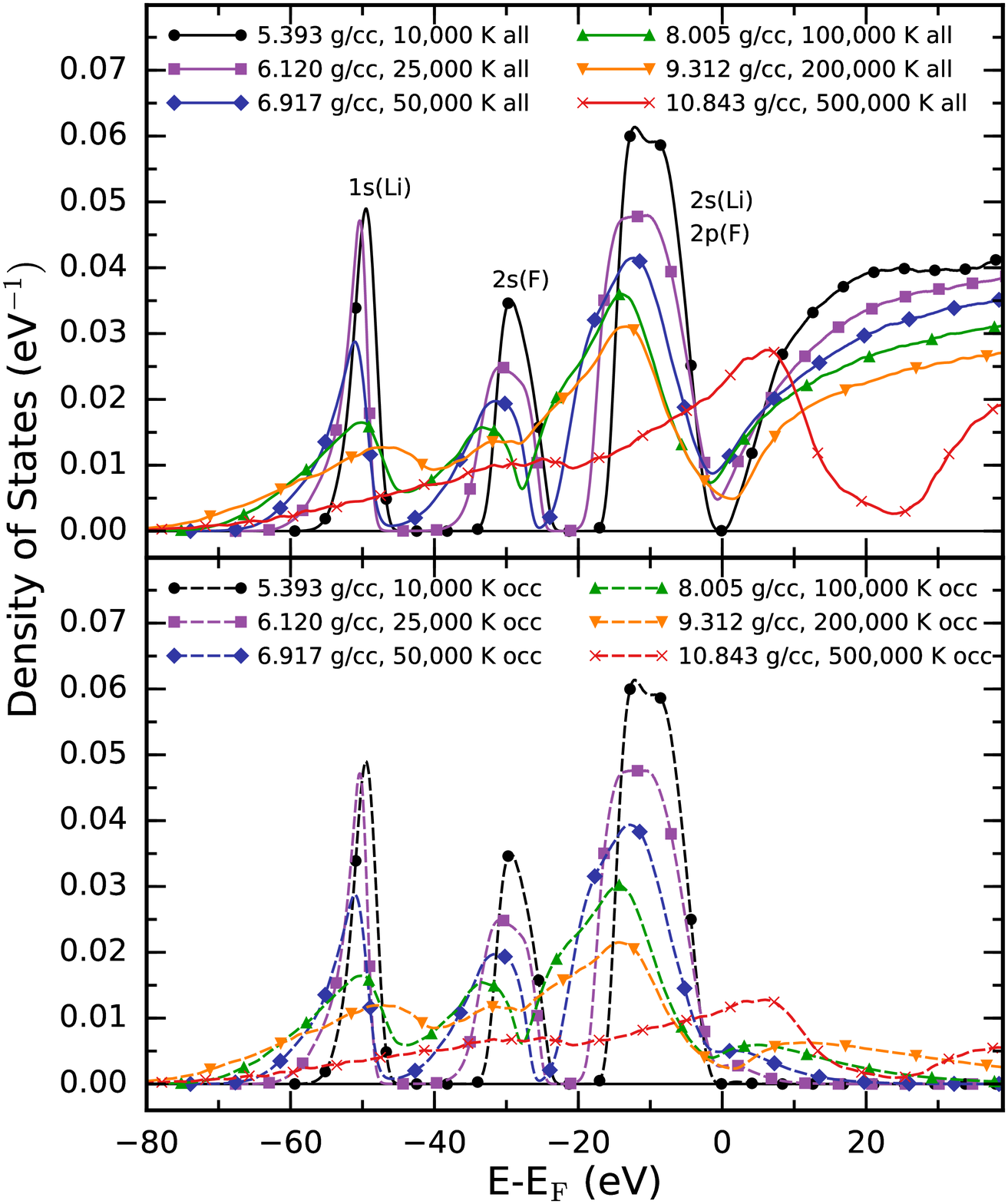}
  \end{center}

    \caption{Electronic DOSs plotted at temperature-density conditions
      along the principal shock Hugoniot curve. The upper panel shows
      the total DOS (all), while the lower panel shows occupied DOS
      (occ).}

  \label{DOS1}
\end{figure}

Fig.~\ref{DOS1} shows the total and occupied DOS at six points along
the liquid and plasma shock Hugoniot curve as predicted by DFT-MD (see
Section VI). At low temperatures and densities, the general structure
is composed of three peaks below the Fermi energy, representing the
atomic 1s, 2s and 2p states. The 1s peak is due to lithium, which is
treated with an all-electron pseudopotential, while the fluorine
pseudopotential has a frozen 1s core.  Depending on the
temperature-density conditions, the DOS exhibits a gap or pseudogap
followed by a continuous spectrum of conducting states. The lowest
density-temperature condition exhibits a gap, which is consistent with
the work of Cl\'{e}rouin \emph{et al.}~\cite{Clerouin2005}, who showed
a gap in the liquid persists along the Hugoniot curve to the melting
point over density range of 4.5--6.5 g$~$cm$^{-3}$. The DOS at higher
temperature-density conditions exhibits a pseudogap, whose depth
generally decreases for increasing temperature-density conditions
along the Hugoniot. The DOS peaks broaden and merge at higher
temperatures and densities as LiF becomes further ionized. For the
highest temperature and density, the total DOS begins to resemble that
of an ideal fluid.

Regarding the occupied DOS, the fraction of occupied states lying
above the Fermi energy drastically increases as temperature-density
conditions increase along the Hugoniot curve. Consistent with our
pair-correlation analyses, we attribute the increase in occupation
above the Fermi energy to a combination of thermal and pressure
ionization. For lithium ions, both L-shell and K-shell states undergo
a significant thermal ionization, while for fluorine ions, only the
L-shell states are subject to thermal ionization in the temperature
range considered for the DOS. The L-shell states in both lithium and
fluorine are partially pressure ionized at the highest densities
considered here, but 1s states remain bound for the conditions
considered in the DOS plot.

\begin{figure}[!htbp]
  \begin{center}
        \includegraphics*[width=8.6cm]{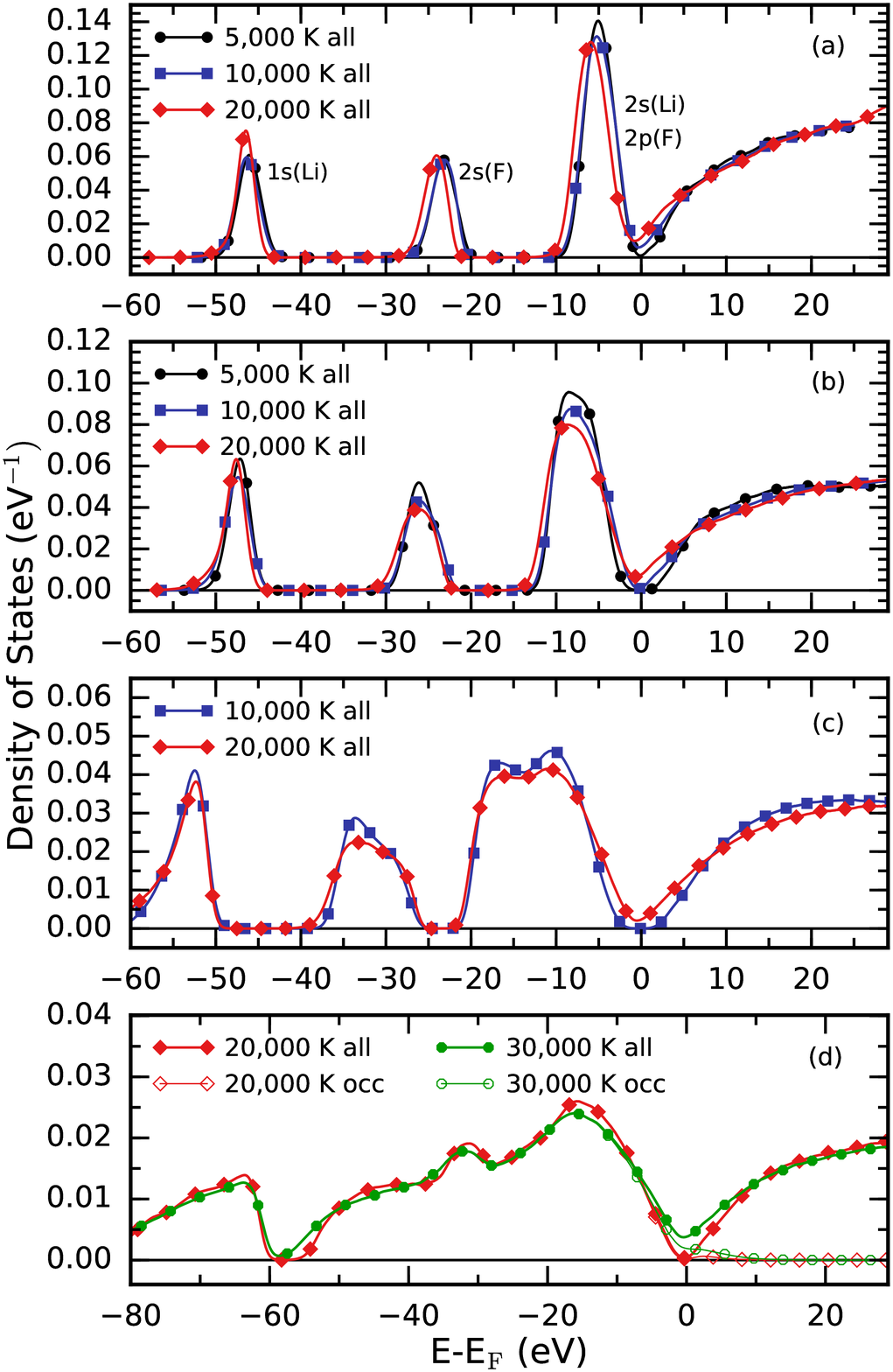}
  \end{center}

    \caption{Electronic DOSs at fixed, liquid temperatures, compared
      for densities of (a) 2.082, (b) 3.651, (c) 7.582, and (d) 15.701
      g$\,$cm$^{-3}$.}

  \label{DOS2}
\end{figure}

Fig.~\ref{DOS2} shows a set of DOS curves for a fixed temperatures as
a function of density for off-Hugoniot states lying in the
low-temperature liquid regime. While all curves shown are found to be
liquid in the DFT-MD simulations, we note that any lower temperatures
than those shown resulted in a frozen structure using 64-atom
simulation cells. In this temperature-density regime, we find that the
band gap forms at increasingly higher temperatures with increasing
density. Based on our ionic pair correlation analysis in
Fig.~\ref{GofRLowT}, we find the reason for this trend in the band gap
is due to ordering maintained within the fluid due ionic Coulomb
interactions. While higher temperatures tend to disorder the fluid,
closing the gap, higher densities stabilize an ionic structure that
promotes a gap.  We also note that, for the low temperatures and high
densities, it is clear that the K-shell states are not pressure
ionized even at the highest density studied here (15.7 g$~$cm$^{-3}$)
and the majority of occupied states still lie below the Fermi energy.

\section{CONCLUSIONS}

In this paper, we have extended the first-principles EOS of LiF to a
much wider temperature-density range than it has ever been studied
previously. For the first time, we are able to predict the compression
maximum on the principal Hugoniot from first principles. We used PIMC
and DFT-MD to construct a coherent EOS that bridges the liquid, WDM,
and plasma regimes. We showed that both PIMC and DFT-MD produce
consistent EOS data in the range of
5$\times$10$^5$--1$\times$10$^6$~K, validating the use of
free-particle nodes in PIMC and zero-temperature XC functionals in
DFT-MD for warm dense LiF. We then studied pair-correlations of
electron and nuclei in LiF liquid and plasmas, revealing an evolving
plasma structure and ionization process that is driven by thermal and
pressure ionization effects. In addition, we computed the density of
states to show how LiF can maintain an open band gap to densities as
high as 15 g$~$cm$^{-3}$ due to strong ionic correlations. Finally, we
examined the shock compression behavior of LiF and computed a
first-principles benchmark of the principal Hugoniot for several
pre-compression conditions. We compare our PIMC Hugoniot results with
widely used Thomas-Fermi-based models (SESAME and LEOS), which do not
include shell effects, and a DFT-based average-atom Purgatorio model,
which agrees well with PIMC, but is slightly stiffer. Overall, we
demonstrate that PIMC is an important tool to benchmark the EOS in the
WDM regime.  Kohn-Sham based DFT simulations are too inefficient to
access physics at temperatures corresponding to the core ionization,
and more efficient, but approximate models do not necessarily capture
all of the complex physics of the WDM regime.

\begin{acknowledgments}

This research is supported by the U. S. Department of Energy, grants
DE-SC0010517 and DE-SC0016248. This research used resources of the
National Energy Research Scientific Computing Center, a DOE Office of
Science User Facility supported by the Office of Science of the
U.S. Department of Energy under Contract No. DE-AC02-05CH11231. This
research also used the Janus supercomputer, which is supported by the
National Science Foundation (Grant No. CNS-0821794) at the University
of Colorado and the National Center for Atmospheric Research.  This
research is also part of the Blue Waters sustained-petascale computing
project (NSF ACI 1640776), which is supported by the National Science
Foundation (awards OCI-0725070 and ACI-1238993) and the state of
Illinois. Blue Waters is a joint effort of the University of Illinois
at Urbana-Champaign and its National Center for Supercomputing
Applications.  We would like to thank Heather Whitley and Christine Wu
for help obtaining the LEOS and Purgatorio[Lynx] EOS models. We would
like to thank Scott Crockett and Sven Rudin for obtaining the SESAME
EOS.

\end{acknowledgments}



\end{document}